\documentclass[runningheads,a4paper]{llncs}
\usepackage{makeidx}
\usepackage{makecell}
\usepackage{booktabs}
\usepackage{mathtools}
\usepackage{footnote}
\usepackage{nccmath}
\usepackage{amsmath}
\usepackage{amsfonts}
\usepackage[utf8]{inputenc}
\usepackage{acronym}
\usepackage{geometry}
\usepackage[hidelinks]{hyperref}
\usepackage{float}
\usepackage{graphicx}
\usepackage{cite}
\usepackage{verbatimbox}
\graphicspath{ {Figures/}{imgs/} }

\usepackage[subtle]{savetrees}
\usepackage{microtype}

\begin{document}

\frontmatter

\title{MinCall~---~MinION end2end convolutional deep learning basecaller}
\author{Neven Miculinić \and Marko Ratković \and Mile Šikić\\
\texttt{\{neven.miculinic, marko.ratkovic, mile.sikic\}@fer.hr}}
\authorrunning{Neven Miculinić \and Marko Ratković \and Mile Šikić}
\institute{University of Zagreb, Faculty of
Electrical Engineering and Computing, Zagreb, Croatia}
\maketitle
% \tableofcontents

\begin{abstract}
    The Oxford Nanopore Technologies's MinION is the first portable DNA sequencing device. It is capable of producing long reads, over 100 kBp were reported. However, it has significantly higher error rate than other methods.
    In this study, we present MinCall, an end2end basecaller model for the MinION. The model is based on deep learning and uses convolutional neural networks (CNN) in its implementation. For extra performance, it uses cutting edge deep learning techniques and architectures, batch normalization and Connectionist Temporal Classification (CTC) loss.
% * <kkrizanovic@gmail.com> 2017-07-03T08:00:15.606Z:
% 
% > For extra performances
% To improve performance, nikako performances
% 
% ^ <kkrizanovic@gmail.com> 2017-07-03T08:21:44.185Z.
    The best performing deep learning model achieves 91.4\% median match rate on \textit{E. Coli} dataset using R9 pore chemistry and 1D reads.

\textbf{Availability:} MinCall is available at \url{https://github.com/nmiculinic/minion-basecaller/} under the MIT license.
% * <marko.ratkovic93@gmail.com> 2017-07-02T18:56:33.288Z:
% 
% Keep keywords?
% 
% ^ <marko.ratkovic93@gmail.com> 2017-07-03T11:16:58.010Z.
\end{abstract}
% * <marko.ratkovic93@gmail.com> 2017-07-02T13:22:38.463Z:
% 
% #ASKMILE
% How to order references by appearance in tex? Is that needed?
% #ODGOVOR
% Sort them in order of first appearance
% 
% 
% ^ <marko.ratkovic93@gmail.com> 2017-07-02T15:36:56.010Z:
% 
% Springer template keeps them in alphabetical order
% I resolved that using https://tex.stackexchange.com/questions/175219/how-to-change-a-bst-for-references-sorted-in-order-of-appearance
% commenting out one line in .bst file 
%
% ^ <mile.sikic@gmail.com> 2017-07-02T15:44:55.960Z:
% 
% great!!
%
% ^ <marko.ratkovic93@gmail.com> 2017-07-03T11:17:01.796Z.
\section{Introduction}
In recent years, deep learning methods significantly improved the state-of-the-art in multiple domains such as computer vision, speech recognition, and natural language processing \cite{LeCun:1998:CNI:303568.303704, NIPS2012_4824}.
In this paper, we present application of deep learning for DNA basecalling problem.

Oxford Nanopore Technology's MinION nanopore sequencing platform~\cite{mikheyev2014first} is the first portable DNA sequencing device. It produces longer reads than competing technologies. In addition, it enables real-time data analysis which makes it suitable for various applications.
% * <kkrizanovic@gmail.com> 2017-07-03T08:02:39.202Z:
% 
% > that make it
% Which makes it suitable
% 
% ^ <kkrizanovic@gmail.com> 2017-07-03T08:22:10.579Z.

Although MinION is able to produce long reads, even up to 882 kb~\cite{loman1-100k,loman2-800k}, they have an error rate of 10\% or higher. This has been somewhat alleviated with new R9 pore model, which replaced previous R7 one. In this paper, we show that this error rate can be reduced by our approach with the properly trained neural network model.
% * <kkrizanovic@gmail.com> 2017-07-03T08:04:36.179Z:
% 
% >  replacing older R7 ones
% Obrisati
% 
% ^ <kkrizanovic@gmail.com> 2017-07-03T08:22:24.998Z.
% * <kkrizanovic@gmail.com> 2017-07-03T08:03:09.407Z:
% 
% > above
% or higher
% 
% ^ <kkrizanovic@gmail.com> 2017-07-03T08:22:27.545Z.

\subsection{Sequencing overview}
Conceptually, the MinION sequencer is a variation on the now standard shotgun sequencing approach. First, DNA is sheared into smaller fragments and adapters are ligated to either end of the fragments. The resulting DNA fragments pass through a protein embedded in a membrane via a nanometre-sized channel, a nanopore. A single DNA strand passes through the pore. Optionally, hairpin protein adapter can connect two DNA strands, allowing both template and complement read passing through the nanopore sequentially for more accurate reads. This technique is referred to as 2D reads. However or focus is on 1D reads containing only single-strand DNA and no hairpin adapter.
% * <kkrizanovic@gmail.com> 2017-07-03T08:25:15.631Z:
% 
% > This technique is referred to as 2D reads. However or focus is on 1D reads containing only template DNA and no hairpin adapter.
% Is it really only template DNA and no complement, Maybe say single-strand?
% 
% ^ <marko.ratkovic93@gmail.com> 2017-07-03T11:17:47.541Z:
% 
% I agree
%
% ^ <marko.ratkovic93@gmail.com> 2017-07-03T12:27:31.606Z.
As DNA strand passes through the nanopore, they are propelled by the current. However, this current varies depending on specific nucleotide context within the nanopore, changing its resistance. Currency is sampled multiple times per second, 4000Hz in our dataset, and from this data, passing DNA fragment is deduced. By design, pores are 6 nucleotides wide, and many models use this information internally. However, we created a model independent of pore width which requires less feature engineering.
% * <kkrizanovic@gmail.com> 2017-07-03T08:27:08.645Z:
% 
% > opted ours out of it
% Not a formal expression so I wouldn't use it in a paper. It also doesn't seem to be correctly used. I would say something like: "so we decided no to use it in our model"
% 
% ^ <marko.ratkovic93@gmail.com> 2017-07-03T11:36:51.682Z:
% 
% I agree. Updated
%
% ^ <mile.sikic@gmail.com> 2017-07-03T13:03:17.958Z.
% * <marko.ratkovic93@gmail.com> 2017-07-02T12:02:20.669Z:
% 
% quote needed
% 
% ^ <marko.ratkovic93@gmail.com> 2017-07-03T11:19:10.330Z.

% \begin{figure}[!ht]
%     \begin{center}
%         \includegraphics[width=0.6\textwidth]{./imgs/sequencing.png}
%         \caption{Depiction of shotgun sequencing}
%         \label{fg:sequencing}
%     \end{center}
% \end{figure}

% \begin{figure}[!ht]
%     \begin{center}
%         \includegraphics[width=0.7\textwidth]{imgs/nanopore.png}
%         \caption[DNA strain being pulled through a nanopore]{DNA strain being pulled through a nanopore \protect\footnotemark}
%         \label{fg:nanopore}
%     \end{center}
% \end{figure}
% \footnotetext{Figure adapted from https://nanoporetech.com/how-it-works}
% * <marko.ratkovic93@gmail.com> 2017-07-02T12:03:39.111Z:
% 
% Image not needed, remove to save space
% 
% ^ <neven.miculinic@gmail.com> 2017-07-02T13:08:12.057Z:
% 
% lgtm
%
% ^ <marko.ratkovic93@gmail.com> 2017-07-02T17:30:30.371Z.

\subsection{Related work}
The core of the decoding process is the basecalling step, that is translating the current samples to the nucleotide sequence. Nowadays there are multiple basecalling options, both official and unofficial ones.

Earlier models were Hidden Markov model (HMM)-based where hidden state modeled DNA sequence of length 6 (6-mer) in the nanopore. Pore models were used in computing emission probabilities.~\cite{loman2015complete,schreiber2015analysis,szalay2015novo,timp2012dna} and the recent open source HMM-based basecaller Nanocall~\cite{david2016nanocall}. Modern basecallers use RNN base models, and in this paper, we used CNN with beam search instead.
% * <kkrizanovic@gmail.com> 2017-07-03T08:46:05.651Z:
% 
%  opted out
% 
% 
% ^ <marko.ratkovic93@gmail.com> 2017-07-03T11:36:32.512Z.

We compared our model on R9 chemistry with Metrichor (HMM-based approach), Nanonet\footnote{\url{https://github.com/nanoporetech/nanonet/}} and DeepNano~\cite{deepnano} (RNN based approaches). Detailed basecaller overview can be found in \cite{denovo_minion}.

\section{Dataset}

Used dataset E.Coli K-12 strands from~\cite{loman1-100k} has been previously passed through MinKNOW and has been basecalled by Metrichor. Since the focus of this paper was 1D read analysis, only 1D reads were used.

\begin{savenotes}
    \begin{table}[htb]
        \caption{Used dataset}
        \label{tbl:datasets}
        \centering

        \begin{tabular}{lcc| c}
            \toprule
            {} &  \thead{Number of reads} &   \thead{Total bases \lbrack bp\rbrack\footnote{Total number of bases called by Metrichor}} &    \thead{Whole genome size \lbrack bp\rbrack} \\
            \midrule
            \emph{{E. Coli}} & 164471 & 1 481 687 490 & 4 639 675\\
%             \footnote{R9 sequencing data from \url{http://lab.loman.net/2016/07/30/nanopore-r9-data-release/}, reference taken from \url{https://www.ncbi.nlm.nih.gov/nuccore/48994873}} 

            \bottomrule
        \end{tabular}
    \end{table}
\end{savenotes}
% * <marko.ratkovic93@gmail.com> 2017-07-02T13:25:21.296Z:
% 
% lambda removed completely
% 
% ^ <marko.ratkovic93@gmail.com> 2017-07-02T17:48:15.827Z.
\subsection{Data preprocessing}

To help the training process, raw signal is split into smaller blocks that are used as inputs. For each Metrichor basecalled event it is easy to determine the block it falls into using \emph{start} field. Using this information output given by Metrichor can be determined for each block.

To correct errors produced by Metrichor and possibly increase the quality of data, each read is aligned to the reference. This is done using aligner GraphMap~\cite{sovic2016fast} that returns the best position in the genome, which is hopefully, the part of the genome from which the read originated.

\begin{figure}[!ht]
    \begin{center}
        \includegraphics[width=1\textwidth]{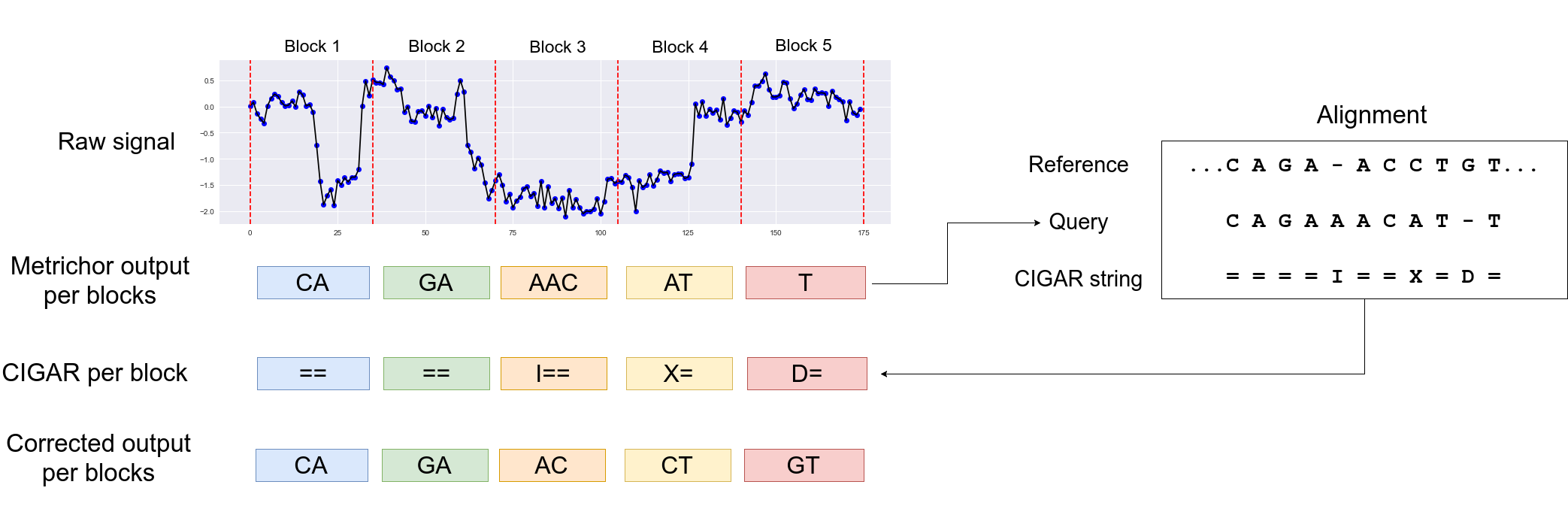}
        \caption{Dataset preparation}
        \label{fg:data_correction}
    \end{center}
\end{figure}

Alignment part in the genome is used as a target. Using CIGAR string returned by aligner we can correct Metrichor data and get target output for each block. This process is shown in Fig. ~\ref{fg:data_correction}.

To eliminate the possibility of overfitting to the known reference, the model is trained and tested on reads from different sources. Due to the limited amount of public available raw nanopore sequence data, E. Coli was \emph{divided} into two regions.
Reads were split into train and test portions, depending on which region of E. Coli they align.
% * <kkrizanovic@gmail.com> 2017-07-03T08:50:50.975Z:
% 
% > Reads were split into train and test portions, depending on which region of E. Coli they align.
% > If read aligns inside first 70\% of the E. Coli, it is placed into train set, and if it aligns to the second portion, it is placed into test set. Reads whose alignment overlaps train and test region are not used. Important to note that E. Coli genome is cyclical, so reads with alignments that wrap over edges are also discarded. Total train set consist of over 110 thousand reads.
% 
% I'm not sure if it is correct to split the whole genome into two parts. Why not only split the dataset?
% 
% ^ <marko.ratkovic93@gmail.com> 2017-07-03T11:29:20.091Z:
% 
% Dataset is split, depending where reads align on the reference.  
% As we are using alignment information to correct read data used during training, we don't want our model to overfit to specific parts of E. Coli, with this model is tested on reads that align to completely different part of E. Coli without leakage of information between train and test.
% 
% ^ <mile.sikic@gmail.com> 2017-07-03T12:36:51.382Z.
If read aligns inside first 70\% of the E. Coli, it is placed into train set, and if it aligns to the second portion, it is placed into test set. Reads whose alignment overlaps train and test region are not used. Important to note that E. Coli genome is cyclical, so reads with alignments that wrap over edges are also discarded. Total train set consist of over 110 thousand reads.

% Overview of the entire learning pipeline is shown in figure~\ref{fg:train_pipe}.
% \begin{figure}[!ht]
%     \begin{center}
%         \includegraphics[width=0.7\textwidth]{./imgs/train_pipeline.png}
%         \caption{Training pipeline overview}
%         \label{fg:train_pipe}
%     \end{center}
% \end{figure}

\begin{verbbox}
    Target      :  A  G  A  A  A  A  A  A  A
    Preprocessed:  A  G  A  A' A  A' A  A' A
\end{verbbox}

\begin{figure}[!htb]
    \centering
    \theverbbox
    \caption{Target nucleotide sequence preprocessing}
    \label{fig:data_preprocessing}
\end{figure}

Due to CTC merged nature during decoding, adjacent duplicates are merged into one, we preprocess the target nucleotide sequence with surrogate nucleotides, such that each second repeated nucleotide is replaced with its surrogate. The example is provided in Fig.~\ref{fig:data_preprocessing}. All raw input data were normalized to zero mean and unit variance as it yields superior performance with neural networks.

\section{Method}
Instead of opting for the traditional path using HMM or newly adopted RNN we tried using residual CNN~(Convolutional neural networks)~\cite{he2016deep}. For loss, we used CTC (Connectionist Temporal Classification)~\cite{graves2006connectionist} between basecalled and target sequence. Other building blocks used are Batch normalization (BN)~\cite{BNORM} and pooling layers. %Dropout~\cite{srivastava2014dropout} was not used.
Described model is implemented using tensorflow~\cite{tensorflow2015-whitepaper} and open source warp-ctc~\cite{warpctc} GPU CTC loss implementation.

The final model is a residual neural network consisting of 72 residual blocks BN\footnote{Batch normalization}-ELU\footnote{Exponential Linear Unit}
%     \begin{equation*}
%     \text{ELU}(x)=
%     \begin{cases}
%     x, & \text{if}\ x>0 \\
%     \alpha (exp(x) - 1), & \text{otherwise}
%     \end{cases}    \\
%     \end{equation*}
-CONV\footnote{1D convolutional layer}-BN-ELU-CONV, to a grand total of 2 million parameters. The used model is a variant of architecture proposed in paper~\cite{identitet} with the difference of ELU being used as activation instead of ReLU as it is reported \cite{resnet-elu} to speeds up the learning process and improve accuracy as the depth increases.

Each convolutional layer in this models uses 64 channels with kernel size 3. Because sequenced read is always shorter than the raw signal, pooling with kernel size two is used every 24 layers resulting in a reduction of dimensionality by factor 8. This is beneficial in faster learning, better generalization and increased basecalling speed.

Training the model is the minimization of previously described CTC loss. It was done using Adam~\cite{adam} with default parameters, exponentially decaying learning rate starting from $10^{-3}$, decay rate of $5\cdot 10^{-2}$ over 100k steps\footnote{We use tf.train.exponential\_decay where current learning rate, lr is $lr=\text{initial\_lr} \cdot \text{decay\_rate}^\frac{\text{global\_step}}{\text{decay\_step}}$} and minibatch size 8. To prevent gradients exploding on bad inputs, they were clipped to range [-2, 2]. We observed no overfitting due to large dataset size.

During development, we tried ReLu and PrELU~\cite{prelu} with no significant result difference. Also, different channel numbers, receptive field width, and various other hyperparameters have been tested  during hyperparameter optimization. SigOpt~\cite{dewancker2016strategy}\footnote{\url{https://sigopt.com/}} bayesian hyperparameter optimization library was used.
% * <kkrizanovic@gmail.com> 2017-07-03T08:56:29.901Z:
% 
% > and conclude after enough complexity, that is sufficient layers, all choices were performing similarly
% I do not understad what is this supposed to mean.
% 
% ^ <marko.ratkovic93@gmail.com> 2017-07-03T11:36:00.700Z:
% 
% @Neven?
%
% ^ <marko.ratkovic93@gmail.com> 2017-07-03T11:49:36.565Z:
% 
% I vote that we skip that part and just mention
% "Also, different channel numbers, receptive field width, and various other hyperparameters during hyperparameter optimization have been tested."
%
% ^ <mile.sikic@gmail.com> 2017-07-03T12:40:55.231Z.

\section{Results}

Developed tool was compared with other available basecallers that support R9 chemistry. This includes third-party basecaller DeepNano and official basecallers by Oxford Nanopore (cloud-based Metrichor and Nanonet).

The exact error rate metric is unreliable since multiple pipeline tools could be the issue. First the sample is prepared, hopefully, uncontaminated and matching reference genome as close as possible then sequenced using the MinION device obtaining raw data. Next, our model (or any other) is applied to basecall the sequences. To evaluate error rate metric basecalled read is aligned to the reference genome using  GraphMap~\cite{sovic2016fast}.
% * <marko.ratkovic93@gmail.com> 2017-07-02T12:42:59.019Z:
% 
% Remove cite for BWA-MEM, GraphMap, sigopt and leave only footnote?
% 
% ^ <neven.miculinic@gmail.com> 2017-07-02T13:09:52.749Z:
% 
% Instead let's skip BWM-MEM entires, mention only Graphmap. I'd leave citation for Sigopt
%
% ^ <marko.ratkovic93@gmail.com> 2017-07-02T13:12:01.531Z:
% 
% Removed part with the bias, left only mention to GraphMap
%
% ^ <marko.ratkovic93@gmail.com> 2017-07-02T17:30:39.260Z.

The fact that ground truth is not known makes evaluation difficult. Different methods for evaluation were used to get clearer information about each basecaller. Different evaluation metrics are described and analyzed in the following subsections. For completeness, speed measurements are given in Table~\ref{tbl:speeds}.

% \begin{table}[htbp]
% 	\caption{Base calling speeds measured in \textit{base pairs per second}. CPU is Intel Xeon E5-2640 v2 @ 2Ghz. GPU is NVIDIA Titan X Black}
% % * <mile.sikic@gmail.com> 2017-07-02T15:53:28.349Z:
% % 
% % Why do  not present results in two columns. This will spare some space :)
% % 
% % ^ <marko.ratkovic93@gmail.com> 2017-07-02T16:10:44.316Z:
% % 
% % Originally there were two columns (ecoli and lambda). 
% % I'll try to transpose it
% %
% % ^.
% 	\label{tbl:speeds}
% 	\centering

% 	\begin{tabular}{lcc}
% 		\toprule
% 		{} &  \thead{Speed (bp/s)}  \\
% 		\midrule
% 		MinCall (CPU)  &               \textbf{ 1363.340 }\\
% 		Nanonet (CPU)  &                897.499 \\
% 		DeepNano (CPU) &                 692.370 \\
% 		\midrule
% 		MinCall (GPU)  &              \textbf{ 6571.76 } \\
% 		Nanonet (GPU)  &               3828.39  \\
% 		\bottomrule
% 	\end{tabular}

% \end{table}

% * <mile.sikic@gmail.com> 2017-07-02T15:53:28.349Z:
% 
% Why do  not present results in two columns. This will spare some space :)
% 
% ^ <marko.ratkovic93@gmail.com> 2017-07-02T16:10:44.316Z:
% 
% Originally there were two columns (ecoli and lambda). 
% I'll try to transpose it
%
% ^ <marko.ratkovic93@gmail.com> 2017-07-02T17:30:43.701Z.
\begin{table}[htbp]
% * <mile.sikic@gmail.com> 2017-07-02T16:20:41.516Z:
% 
% Table 2 and three should be the same. 
% Rows (Tools, Speed CPU(bp/s), Speed GPU(bp/s) )
% Columns (Row names, MinCall, Nanonet, DeepNano)
% Put in the caption explanation about the DeepNano and GPU - it does not support or something
% 
% ^ <marko.ratkovic93@gmail.com> 2017-07-02T16:29:31.447Z:
% 
% Updated
%
% ^ <mile.sikic@gmail.com> 2017-07-02T16:37:07.177Z:
% 
% Please double check. CPU is faster than GPU ?
%
% ^ <marko.ratkovic93@gmail.com> 2017-07-02T17:19:55.676Z:
% 
% Am I missing something?  GPU 6571.76  vs CPU 1363.34?
% 
% ^ <mile.sikic@gmail.com> 2017-07-02T17:37:15.626Z:
% 
% forget, my mistake :)
%
% ^ <marko.ratkovic93@gmail.com> 2017-07-02T17:49:00.586Z.
	\caption{Base calling speeds measured in \textit{base pairs per second}. Tested on Intel Xeon E5-2640 v2 @ 2Ghz with NVIDIA Titan X Black GPU. DeepNano does not support GPU basecalling.}
	\label{tbl:speeds}
	\centering
	\begin{tabular}{lccc}
		\toprule
		{} &  \thead{MinCall} &  \thead{Nanonet} &  \thead{DeepNano}  \\
		\midrule
		Speed CPU(bp/s)  & \textbf{ 1363.34 } & 897.49 & 692.37 \\
		Speed GPU(bp/s)  & \textbf{ 6571.76 } & 3828.39  & - \\
		\bottomrule
	\end{tabular}

\end{table}

\subsection{Per read metrics}
\label{subs:read_metrics}
\begin{figure}[htb]
    \begin{center}
        \includegraphics[width=0.8\textwidth]{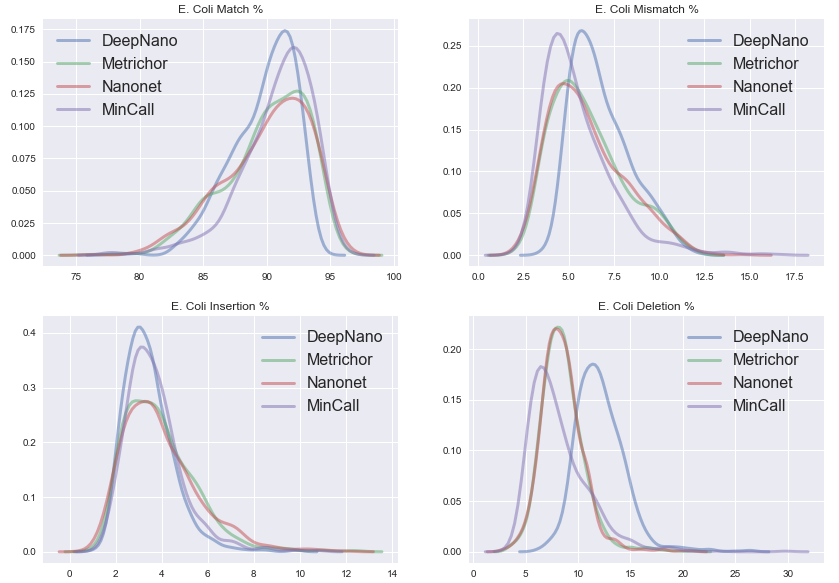}
        \caption{KDE plot for the distribution of percentage of alimnment operations for E. Coli}
        \label{fg:ecoli_kde}
    \end{center}
\end{figure}

A portion of the read length that aligns correctly is called match\_rate. Same goes for mismatches and insertions.
The sum of all matches, mismatches, and insertions is equal to the read length. Results on E.Coli test set with Graphmap aligner are shown in Table~\ref{tbl:ecoli_rates}. Furthermore, we plot Kernel Density Estimation(KDE) plots for each mentioned statistic on E. Coli dataset in Fig. ~\ref{fg:ecoli_kde}.

% Analysis of how matches, mismatches, insertions, and deletions are distributed across the reads has shown that mismatches and insertion occur more frequently at the beginnings and the ends of the reads. This is not only the case for the developed basecaller, but all other shows the same property. This could be due to lack of context information from both sides when edges are basecalled.
% * <marko.ratkovic93@gmail.com> 2017-07-02T12:26:24.877Z:
% 
% Removed mention of Appendix.
% Maybe leave like this, just the mention or completely remove this?
% 
% ^ <neven.miculinic@gmail.com> 2017-07-02T13:10:41.468Z:
% 
% I vote remove completely
%
% ^ <marko.ratkovic93@gmail.com> 2017-07-02T13:13:16.638Z:
% 
% Commented-out that part
% #ASKMILE
% 
% ^ <mile.sikic@gmail.com> 2017-07-02T16:17:32.028Z:
% 
% Remove everything connected to Appendix (Btw, Sebrek has had appendix surgery tonight). When we have results on other datasets we will write a full paper with all details
% 
% ^ <marko.ratkovic93@gmail.com> 2017-07-02T17:30:49.566Z.

\begin{table}[htbp]
    \caption{Alignment specifications of E. Coli R9 basecalled reads using GraphMap}
    \label{tbl:ecoli_rates}
    \centering
    \begin{tabular}{lcccc}
        \toprule
        {} &  \thead{Match \% \\(median)} &  \thead{Mismatch \% \\(median)} &  \thead{Insertion \% \\(median)} &  \thead{Deletion \% \\(median)} \\
        \midrule
        DeepNano   &                  90.254762 &                      6.452852 &                       \textbf{3.274420} &                     11.829965 \\
        Metrichor  &                  90.560455 &                      5.688105 &                       3.660381 &                      8.328271 \\
        Nanonet    &                  90.607674 &                      5.608912 &                       3.652791 &                      8.299046 \\
        MinCall    &                  \textbf{91.408591} &                     \textbf{ 5.019141} &                       3.477739 &                      \textbf{7.471608 }\\
        \bottomrule
    \end{tabular}
\end{table}

\subsection{Consensus metrics}

In Subsection~\ref{subs:read_metrics} we showed metrics after the read is aligned to reference genome. In a second approach, we reconstruct the original genome from basecalled reads. 

\subsubsection{Consensus from pileup}

\begin{figure}[!htb]
   \begin{center}
       \includegraphics[width=0.8\textwidth]{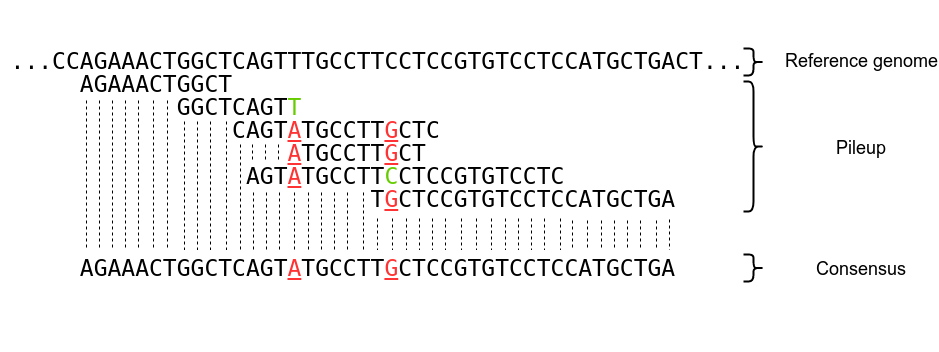}
       \caption{Consensus from pileup}
       \label{fg:consensus}
   \end{center}
\end{figure}

Since the reference genome of E. Coli is known, we simply align all the reads to the genome, stack them on top of each other forming pileup of read bases. Using majority vote, dominant bases are called on each position. The resulting sequence is called consensus. When calling consensus for deletions,  there has to be a majority of deletions of the same length. Calling insertions has an additional condition, the majority has to agree on both length and the bases of insertion. Fig. ~\ref{fg:consensus} shows how consensus is called from pileup created from aligned reads. Pileup is stored in mpileup format.

All models show a slight bias towards deletions than insertions, but this may be the limitation of technology as it has been reported that deletion and mismatch rates for nanopore data are ordinarily higher than insertion rates~\cite{sovic2016fast}. Results are shown in Table~\ref{tbl:spec_ecoli}. %Lambda table is deferred to Appendix in table~\ref{tbl:spec_lambda}.

\begin{table}[!htb]
    \caption{Consensus specifications of E. Coli R9 basecalled reads}
    \label{tbl:spec_ecoli}
    \centering
    \begin{tabular}{lcccccc}
        \toprule
        {} &  \thead{Total called\\\lbrack bp\rbrack} &  \thead{Correctly called\\\lbrack bp\rbrack} &  \thead{Match\\\%} &  \thead{Snp\\\%} &  \thead{Insertion\\\%} &  \thead{Deletion\\\%} \\
        \midrule
        DeepNano  &                  1510244.0 &                      1493242.0 &          98.8742 &         1.0044 &               0.1214 &              0.9041 \\
        Metrichor &                  1515893.0 &                      1502588.0 &          99.1223 &         0.7464 &               0.1313 &              0.6300 \\
        Nanonet   &                  1414237.0 &                      1385515.0 &          97.9691 &         1.5700 &               0.4609 &              1.5158 \\
        MinCall   &                  1517828.0 &                      1506233.0 &          \textbf{99.2361} &         \textbf{0.6474} &               \textbf{0.1165} &             \textbf{ 0.5510 }\\
        \bottomrule
    \end{tabular}
\end{table}

\subsubsection{Consensus from de novo assembly}

In this evaluation method, the consensus sequence is calculated using de novo genome assembly. For this task, fast and accurate de novo genome assembler \emph{ra}\footnote{\url{https://github.com/rvaser/ra}}~\cite{vaser} was used and obtained consensus sequence is compared to the reference using dnadiff present in the Mumer\footnote{\url{https://github.com/garviz/MUMmer}}.
The length of the reference, consensus sequence, number of contigs and percentages of aligned bases from the reference to the query and vice versa are shown in the Table \ref{tbl:assembly}. Average identity summarizes how closely does the assembled sequence match the reference. This is run on full E. Coli sequence run for 1D template reads ($\sim$160k reads), for our tool, Nanonet and Metrichor. Developed tool has shown an increase in quality of the assembled sequence over Metrichor by offering longer consensus, higher identity percentage, and overall smaller edit distance\footnote{Calculated using \url{https://github.com/isovic/racon/blob/master/scripts/edcontigs.py}}.

\begin{table}[htb]
    \caption{Assembly and consensus results for E. Coli}
    \label{tbl:assembly}
    \centering
\begin{tabular}{lccc}
\toprule
&         Metrichor &           MinCall &          Nanonet \\
\midrule
% \thead{Ref. genome size (bp)} &           4639675 &           4639675 &            4639675 \\
% \thead{Total bases (bp)}      &           4604806 &           \textbf{4614354} &          4600056 \\
% \thead{Contigs [\#]}           &                 1 &                 1 &                1 \\
\thead{Aln. bases ref. (bp)}  &  4639641(100.00\%) &  4639612(100.00\%) &  4639031(99.99\%) \\
 \thead{Aln. bases query (bp)} &  4604787(100.00\%) &  4614351(100.00\%) &  4599745(99.99\%) \\
\thead{Avg. Identity}         &             98.76 &             \textbf{99.06} &            98.47 \\
\thead{Edit distance}         &             60418 &             \textbf{46686 }&            74341 \\
\bottomrule
\end{tabular}
\end{table}
% * <marko.ratkovic93@gmail.com> 2017-07-02T12:51:11.868Z:
% 
% Remove rows #contigs, total bases and reference genome size?
% 
% ^ <neven.miculinic@gmail.com> 2017-07-02T13:12:19.496Z:
% 
% I agree.
%
% ^ <marko.ratkovic93@gmail.com> 2017-07-02T13:13:55.853Z:
% 
% Also commented-out those rows
% #ASKMILE
% 
% ^ <mile.sikic@gmail.com> 2017-07-02T16:00:18.563Z:
% 
% I agree
%
% ^ <marko.ratkovic93@gmail.com> 2017-07-02T17:30:55.125Z.

\section{Conclusion and further work}

In this paper, we used CNN instead of already tried RNN or HMM approaches, which resulted in higher accuracy compared to other existing basecallers.
Unlike HMM and RNN, there's no explicit dependency on previous hidden state, therefore this model is massively parallelizable and more sensible given data nature --- that is we're dealing with signal processing, not heavily context-depended language modeling.

All test are done on data for R9 chemistry, but the developed open source code could easily be adjusted and trained on R9.4 and newest R9.5 data when it becomes publicly available.

Currently, without support for newer sequencing data, this model has limited application. It can be used as a demonstration of a different approach to basecalling which yields promising results.  As newer versions of basecallers by Oxford Nanopore do not offer any support for data sequenced with previous version of chemistries, this tool can be used to re-basecall that data and to improve the quality of reads retrospectively.

Future work includes experiments with recently proposed Scaled exponential linear units (SELU)~\cite{selu} that eliminate the need for normalization techniques such as used batch normalization. Possible improvements of the model include the combination of convolutions and attention mechanism proposed just recently in the paper~\cite{facebook} showing excellent results for tasks of language translation in both speed and accuracy. Another option could be the usage of stacked simple models, such as logistic regression and SVM to predict each nucleotide given raw signal context, similar to our deep learning model pre-CTC layer, and use linear chain CRF for full sequence basecalling.
% * <kkrizanovic@gmail.com> 2017-07-03T11:07:27.317Z:
% 
% >  for tasks of language translation
% Move this part earlier. Exact location depends on the desired meaning.
% 
% ^ <marko.ratkovic93@gmail.com> 2017-07-03T11:35:07.611Z:
% 
% Not sure what you mean?
% Maybe something like: "showing excellent results in both speed and accuracy for/in/on language translation tasks."
% 
% ^ <mile.sikic@gmail.com> 2017-07-03T12:53:15.373Z.

\section{Acknowledgments}

We're grateful to various other people whose code, tools and advice we've used in completing this project: Fran Jurišić, Ana Marija Selak, Ivan Sović, Robert Vaser and Martin Šošić.

%For data, we're thankful to Loman labs~\cite{loman1-100k} for publicly posting their dataset.

During this paper creation, we used Sigopt academic license for hyperparameter optimization. We gratefully acknowledge the support of NVIDIA Corporation with the donation of the Titan Xp GPU used for this research.
This work has been supported in part by Croatian Science Foundation under the 
project UIP-11-2013-7353 "Algorithms for Genome Sequence Analysis". 

% set style - not sure which one is official for papers
% spbasic (springer basic makes sense)
% \bibliographystyle{spbasic}
\bibliographystyle{splncs03}
\bibliography{refs}

\end{document}